\documentclass{article}
\usepackage{spconf,amsmath,graphicx,hyperref}

\usepackage[utf8]{inputenc} 
\usepackage[T1]{fontenc}    
\usepackage{hyperref}       
\usepackage{url}            
\usepackage{booktabs}       
\usepackage{amsfonts}       
\usepackage{nicefrac}       
\usepackage{microtype}      
\usepackage[dvipsnames]{xcolor}
\usepackage{amsmath} 
\usepackage{graphicx}
\usepackage{multirow}
\usepackage{amssymb}
\usepackage{algorithm}
\usepackage{algpseudocode}
\usepackage{wrapfig}
\usepackage{caption} 
\usepackage[numbers]{natbib}

\usepackage{siunitx} 

\usepackage{array}
\usepackage{subcaption}  
\usepackage{xspace}
\usepackage{comment}
\usepackage{tabularx}
\usepackage{tipa}
\usepackage{enumitem}

\newcolumntype{L}[1]{>{\hspace{6pt}}#1<{\hspace{6pt}}} 

\newcommand{\peh}[1]
{{\color{blue}\bgroup\textbf{[peh:}}~#1\textbf{]}\egroup}

\newcommand{\nic}[1]
{{\color{green}\bgroup\textbf{[nic:}}~#1\textbf{]}\egroup}

\newcommand{\yiq}[1]
{{\color{Maroon}\bgroup\textbf{[yiq:}}~#1\textbf{]}\egroup}

\newcommand{\rom}[1]
{{\color{orange}\bgroup\textbf{[rom:}}~#1\textbf{]}\egroup}

\renewcommand{\paragraph}[1]{\vspace{1em}\noindent\textbf{#1} }



\title{Variational Low-Rank Adaptation for Personalized Impaired Speech Recognition}

\name{Niclas Pokel$^{1,2,*}$, Pehuén Moure$^{1,*}$, Roman Boehringer$^{1,\dagger}$, Shih-Chii Liu$^{1,\dagger}$, Yingqiang Gao$^{3,\dagger}$ \thanks{$^{*}$These authors contributed equally. $^{\dagger}$These authors also contributed equally. This work was partially funded by the Swiss National Science Foundation project CA-DNNEdge (208227).}}
      
\address{$^{1}$Institute of Neuroinformatics, University of Zurich and ETH Zurich, Switzerland \\
      $^{2}$School of Computation, Information and Technology, Technical University of Munich, Germany \\
      $^{3}$Department of Computational Linguistics, University of Zurich, Switzerland \\
      \texttt{\{npokel,pehuen,roman,shih\}@ini.ethz.ch, yingqiang.gao@cl.uzh.ch}
}
\begin{document}
\ninept 

\maketitle

\begin{abstract}

   Speech impairments resulting from congenital disorders, such as cerebral palsy, Down syndrome, or Apert syndrome, as well as acquired brain injuries due to stroke, traumatic accidents, or tumors, present major challenges to automatic speech recognition (ASR) systems.  Despite recent advancements, state-of-the-art ASR models like Whisper still struggle with non-normative speech due to limited training data availability and high acoustic variability. Moreover, collecting and annotating non-normative speech is burdensome: speaking is effortful for many affected individuals, while laborious annotation often requires caregivers familiar with the speaker. This work introduces a novel ASR personalization method based on Bayesian Low-rank Adaptation for data-efficient fine-tuning. We validate our method on the English \textit{UA-Speech} dataset and a newly collected German speech dataset, \textit{BF-Sprache}, from an individual with structural speech impairment. Both the dataset and the approach are designed to reflect the challenges of low-resource settings that include individuals with speech impairments. Our method significantly improves ASR accuracy for impaired speech while maintaining data and annotation efficiency, offering a practical path toward inclusive ASR.

\end{abstract}

\section{Introduction}
Despite having intact cognitive and language abilities, many individuals with speech impairments remain effectively voiceless in a world built around spoken communication. Daily interactions, whether answering a question in class, telling a story, or participating in a group play, can become exhausting, frustrating, or simply impossible. For children, this communication barrier not only isolates them socially but also disrupts their emotional and educational development. The burden extends to families, educators and care providers, who must interpret, advocate and adapt, often without the support of reliable assistive tools \cite{page2022communicative, van2023automatic}.

Automatic speech recognition (ASR) systems hold the potential to bridge this gap, but current models are not designed for non-normative speech. Even state-of-the-art models such as Whisper \cite{radford2023robust} and wav2vec \cite{schneider2019wav2vec, baskar2022speaker} degrade significantly  in performance when confronted with atypical articulation, prosodic variation, or inconsistent phoneme production \cite{Rowe2022}. 

These challenges are further aggravated for non-English languages due to the lack of representative data, limited linguistic tools, and the need for language-specific adaptation strategies \cite{baskar2022speaker, tobin2024automatic}. 
German remains under-resourced for non-normative speech, particularly for children, with limited publicly available datasets \cite{Rumberg2022kidsTALC, Halpern2022, Guldimann2024}.



To address these limitations, researchers have employed various strategies. Fine-tuning large pre-trained models has proven effective for disordered speech, with approaches like those of Shor et al. \cite{shor19_interspeech} achieving 60\% reductions in word error rate for patients with amyotrophic lateral sclerosis. However, such methods are often prone to overfitting and inefficient parameter usage \cite{qi23b_interspeech}. 

Parameter-efficient adaptation methods, including lightweight adapters \cite{qi23b_interspeech}, low-rank adaptation (LoRA) modules \cite{hu2022lora}, and hyper-network-based speaker tuning \cite{mullerEberstein2024hypernetworks}, improve speech foundation model fine-tuning. However, these methods often rely on English language backbones and assume explicit knowledge of impairment types or sufficient in-domain data \cite{hermann2023few, tobin2024automatic}.

Bayesian Neural Network (BNN)-based approaches offer promising alternatives for improving ASR robustness in low-data, high-variability settings \cite{moure2024regularized}, and have been used to maintain robustness across continuous adaptation \cite{ebrahimiuncertainty}. BNN approaches may provide key benefits in bridging parameter-efficient fine-tuning methods to the high variability data of impaired speech. Prior works primarily leverage Bayesian low-rank adaptation (LoRA) for efficiency, either via capacity-reducing pruning and quantization \cite{cong2025improvingloravariationallearning, meo2024bayesianlora} or static post-hoc analysis \cite{yang2024bayesian, Seth2025}, which risks underfitting the complex acoustic features of disordered speech. Conversely, our method employs variational inference (VI) as a dynamic training regularizer that retains full capacity while elastically constraining adaptation to the pre-trained weight structure. 
To this end, our work proposes:
\begin{enumerate}[noitemsep, topsep=0pt, partopsep=0pt, parsep=0pt, left=0pt]
\item \textbf{Variational Low-rank Adaptation framework.} We introduce a Bayesian LoRA method called VI LoRA that captures uncertainty during fine-tuning. This enables robust personalization with significantly less data, while maintaining parameter efficiency, crucial for modeling speech with high acoustic variability.


\item \textbf{Data-driven prior estimation.} We develop a prior estimation approach that better captures the multi-modal distribution of layer-wise weight variations in state-of-the-art ASR architectures.

\item \textbf{Cross-lingual evaluation.} We validate our method on English and German datasets spanning a range of speech intelligibility. Our results show substantial improvements, especially for speakers with very low intelligibility, demonstrating the framework's effectiveness in low-resource, cross-lingual settings.

\end{enumerate}
Together, these contributions bridge existing research gaps by enabling a personalized, interpretable, and scalable ASR solution tailored for users with atypical speech in multiple languages.







\section{Methodology}
\label{sec:Methods}
\subsection{ASR Model and Evaluation Metrics}
For all experiments, we used Whisper-Large V3 \cite{radford2023robust} as the backbone of the impaired speech model, evaluating its performance under both zero-shot and supervised fine-tuning settings, with augmentation via semantic re-chaining (i.e., assembling semantically coherent sentence-level utterances from word-level counterparts \cite{pokel25_diss}). We used word error rate (WER) and character error rate (CER) as metrics to evaluate our experiment outcomes. 


\subsection{Bayesian Low-rank Adaptation Framework}

Our approach addresses the challenges of low data availability, overfitting, and over-parameterization when fine-tuning large models \citep{qi23b_interspeech} and enhances the standard low-rank adaptation (LoRA) \citep{hu2022lora}, a prominent parameter-efficient fine-tuning (PEFT) technique \citep{lialin2024scalingscaleupguide}. LoRA adapts a pre-trained weight matrix $W_0 \in \mathbb{R}^{d_{\mathrm{out}} \times d_{\mathrm{in}}}$ by freezing $W_0$ and introducing a trainable low-rank update $\Delta W_0$. This update is parameterized as the product of two smaller matrices, $B \in \mathbb{R}^{d_{\mathrm{out}} \times r}$ and $A \in \mathbb{R}^{r \times d_{\mathrm{in}}}$, where the rank $r \ll \min(d_{\mathrm{in}}, d_{\mathrm{out}})$. The adapted weight matrix then becomes $W_0 + \frac{\alpha}{r}BA$, with $\alpha$ being a scaling factor.

\begin{figure}[htb]
    \centering
    \vspace{-1\baselineskip}
    \includegraphics[width=\linewidth]{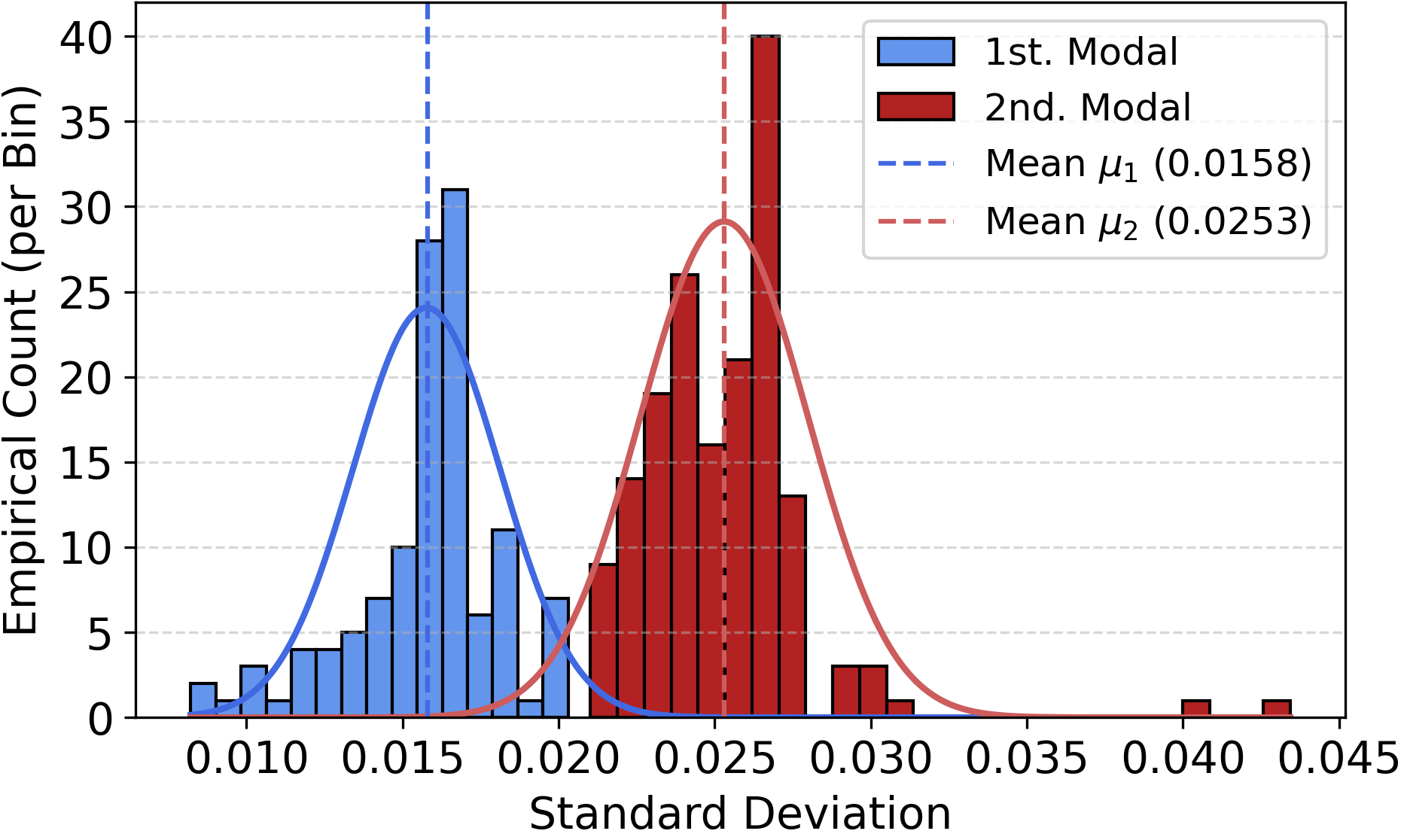}
    \caption{Histogram of the empirically estimated standard deviations, $\hat{\sigma}_{p}^{(l)}$, computed individually for each of the $N=288$ target LoRA layers based on their pre-trained weights $W_0^{(l)}$. Dashed lines indicate the means of the two distinct modes identified by k-means, justifying a layer-type specific prior variance.}
    \label{fig:bimodal}
\end{figure}

While LoRA significantly reduces training efforts, in data-sparse scenarios, the matrices $A$ and $B$ can still overfit the limited training data and cause generalization degradation \citep{lin2024lora}. Drawing inspiration from \textit{Bayes by Backprop} \citep{blundell2015weight}, which regularizes neural networks by learning distributions over their weights, we extend LoRA to a Bayesian setting. This allows us to capture uncertainty about the LoRA parameters, which is particularly beneficial for regularization and improving robustness when training data is scarce. Specifically, we perform variational inference (VI) to estimate the posterior distributions of $A$ and $B$ given the training data $\mathcal{D}$. We approximate the true, often intractable, posterior $p(A,B|\mathcal{D})$ with a tractable variational distribution $q_\phi (A,B)$, parameterized by $\phi$. Adopting the mean-field approximation, we assume independence between $A$ and $B$, and further, between individual elements: $q_{\phi} (A,B) = q_{\phi_A}(A) q_{\phi_B}(B)$, where each factor is a fully diagonal Gaussian:
\begin{align*}
    q_{\phi_A}(A) &= \prod_{i=1}^r \prod_{j=1}^{d_{\mathrm{in}}} \mathcal{N}(A_{ij}| \mu_{A_{ij}}, \sigma^2_{A_{ij}}), \\ 
    q_{\phi_B}(B) &= \prod_{k=1}^{d_{\mathrm{out}}} \prod_{l=1}^r \mathcal{N}(B_{kl}|\mu_{B_{kl}}, \sigma^2_{B_{kl}}).
\end{align*}
The variational parameters $\phi = \{ (\mu_A, \sigma_A), (\mu_B, \sigma_B) \}$ are associated with the LoRA adapter layers (e.g., in query, key, and value projection matrices of multi-head attention) and are learned by minimizing the negative evidence lower bound (ELBO):
\begin{equation*}
\label{alg:elbo}
\resizebox{\columnwidth}{!}{$
\begin{aligned}
    \phi^* & = \arg\min_{\phi} \mathrm{KL} [q_{\phi}(A,B) \,||\, p(A,B|\mathcal{D})] \vphantom{\int} \\ 
    & = \arg\min_{\phi} \int q_{\phi}(A,B) \log \frac{q_\phi(A,B)}{p(A,B)p(\mathcal{D}|A,B)} \, d(A,B) \\
    & = \arg\min_{\phi} \underbrace{\mathrm{KL}[q_\phi(A,B) \,||\, p(A,B)] - \mathbb{E}_{q_\phi(A,B)}[\log p(\mathcal{D} | A,B)]}_{-\mathcal{L}_{\mathrm{ELBO}}(\phi)} \vphantom{\int},
\end{aligned}
$}
\end{equation*}
here, $p(A,B)$ is the prior distribution over the LoRA matrices. The term $\mathbb{E}_{q_\phi(A,B)}[\log p(\mathcal{D} | A,B)]$ is the expected log-likelihood of the training data, corresponding to the task-specific loss (e.g., cross-entropy for ASR) computed with samples from $q_\phi(A,B)$, typically estimated using Monte Carlo sampling.
Let $N$ be the total number of LoRA layers where VI is applied. The KL divergence term in the ELBO, $\mathrm{KL}[q_\phi(A,B) || p(A,B)]$, is theoretically a sum over all $N$ layers. However, numerical instabilities, especially in early training stages before posteriors stabilize, can lead to non-finite KL values for some layers. To ensure robust optimization, our final loss function $\mathcal{L}_{\mathrm{VI}}$ employs an average over layers yielding a finite (non-NaN, non-Inf) KL divergence terms in the current optimization step:
\begin{align*}
    \mathcal{L}_{\mathrm{VI}} = &-\mathbb{E}_{q_\phi(A,B)}[\log p(\mathcal{D}|A,B)] + \\ &\beta \underbrace{\left(\sum \mathrm{KL}[q_\phi (A,B) || p(A,B)] 
    \neq \mathrm{NaN} \text{ or } \mathrm{Inf} \right)}_{\mathrm{\overline{KL}}[q_\phi || p]},
\end{align*}
\noindent
where $\beta$ is a scaling factor. To construct a more informed prior $p(A,B)$, we assume factorization across layers $l$ and between $A^{(l)}$ and $B^{(l)}$, i.e., $p(A,B) = \prod_l p(A^{(l)}) p(B^{(l)})$. For individual elements, we use Gaussian priors: $p(a_{ij}^{(l)}) = \mathcal{N}(a_{ij}^{(l)}|\mu_p, (\sigma_p^{(l)})^2)$ and similarly for $b_{ij}^{(l)}$. A common choice sets $\mu_p=0$ and employs a single, global prior variance, e.g., $\sigma_p^2=1$. However, this assumes that the LoRA updates should operate at a scale comparable to a standard normal distribution, which can be problematic if the pre-trained weights $W_0^{(l)}$ themselves exhibit significantly distinct variances across layers in the pre-trained network parameters, potentially leading to an overly restrictive prior for some layers or an overly loose one for others.
For a more informed prior, we first analyze the empirical standard deviations of pre-trained weights within each target layer $W_0^{(l)}$. The final loss is a weighted sum of the standard Whisper loss (90\%) and a KL divergence term ($10\% \cdot {\text{KL}}(q||p)$), which acts as a regularizer. 
This relative weighting prevents the KL term from dominating when task loss becomes small.
For each layer $l$ designated for LoRA updates, we compute its empirical weight standard deviation $\hat{\sigma}_{p}^{(l)}$ as
\begin{align*}
    \hat{\sigma}_{p}^{(l)} = \sqrt{\frac{1}{|\mathcal{W}^{(l)}| - 1} \sum_{w \in W_0^{(l)}} (w - \overline{w}^{(l)})^2},
\end{align*}
where $\mathcal{W}^{(l)}$ represents the set of all weights in the original matrix $W_0^{(l)}$, and $\overline{w}^{(l)}$ is their mean.
Our empirical analysis of these layer-wise standard deviations, $\{\hat{\sigma}_{p}^{(l)}\}_{l=1}^N$ (across $N=288$ target layers in our experiments), reveals a distinct bimodal distribution, as illustrated in Figure~\ref{fig:bimodal}. A simple Gaussian Mixture Model was used to find the optimal $\mu$ for the layer-specific $\sigma$ prior.

\section{Experimental Setup}
\label{sec:Exper}
We conduct a comprehensive evaluation of our approach across multiple dimensions to assess both personalization effectiveness and generalizability. First, we compare personalization performance against several baselines, including full-parameter fine-tuning, standard low-rank adaptation (LoRA), the high-rank updating technique MoRA \cite{jiang2024mora}, and variational inference LoRA (VI LoRA) with single and bimodal priors. Second, we analyze the impact of architectural design choices on recognition accuracy to understand how specific configurations influence performance. Selecting a comparable rank for MoRA is non-trivial, since the rank must divide the Whisper embedding dimension and cannot be directly matched parameter wise. Moreover, Bayesian methods
and deterministic methods differ fundamentally in parameterization. To align with the parameter count of VI LoRA, including variance terms, we set $r=320$ for MoRA. We also evaluated several (VI-)LoRA ranks ($r$) to optimize the model's configuration. We selected $r=32$ for our experiments, as it provided a strong balance between performance gains on non-normative speech and a minimal increase in catastrophic forgetting. We observed that higher ranks (e.g., $r=64$) failed to further improve target domain performance while exacerbating forgetting, a behavior consistent with the ``intrinsic rank'' hypothesis of LoRA \cite{hu2022lora}.

\subsection{Datasets}
Developing effective ASR systems for individuals with speech impairments requires datasets that capture the acoustic variability, articulatory challenges, and language-specific characteristics of non-normative speech patterns \cite{hustad2019differentiating}.
We evaluated our method using the following datasets: UA-Speech \cite{kim2008dysarthric} and BF-Sprache \cite{pokel25_diss}, as well as Mozilla Common Voice Dataset \cite{ardila-etal-2020-common} for non-impaired speech:  

\paragraph{UA-Speech.} The English UA-Speech dataset \cite{kim2008dysarthric} is widely used in dysarthric speech recognition research and has laid the foundation for numerous deep learning-based works \cite{schu2023using, cadet2024study}. It consists of recordings from 19 speakers with varying degrees of dysarthria, along with recordings from 13 control speakers. We exclusively used the $\approx 66h$ of dysarthic speech. The dataset emphasizes isolated word-level utterances, comprising 455 unique words including digits, letters, and phonetically rich uncommon words. 

\paragraph{BF-Sprache.} To evaluate our personalization framework across languages, we use the German BF-Sprache \cite{pokel25_diss} dataset. For consistency, the dataset also consists of isolated word-level utterances in the training set,  but was tested on spontaneous speech for the most realistic assessment.


The structured design of the two datasets enables controlled experimentation and provides insights into the effectiveness and generalizability of ASR personalization across different intelligibility levels of speech impairment and languages. 


\paragraph{Normative evaluation set.} To measure the forgetting of the already learned distribution of normal speech we used a split in the respective language (German or English) of the widely known Mozilla Common Voice Dataset \cite{ardila-etal-2020-common} for validation and testing.

\subsection{Model Evaluation}
To ensure robust evaluation, we stratify experiments along two primary dimensions:

\begin{enumerate}[noitemsep, topsep=0pt, partopsep=0pt, parsep=0pt, left=0pt]
    \item \textbf{Intelligibility levels.} We partition the UA-Speech dataset into subsets based on reported intelligibility levels (i.e., very low, low, medium), enabling a detailed analysis of model performance across speech impairment levels and the calibration of uncertainty estimates throughout the spectrum.
    
    \item \textbf{Cross-lingual generalization.} We test our methods on both UA-Speech and BF-Sprache datasets to evaluate how well the personalization framework generalizes across languages.
\end{enumerate}

\begin{table}[htb]
\centering
\caption{Results of different models on non-normative (BF-Sprache) and normative (CommonVoice) speech. WD refers to Weight Decay, DP to Dual Prior, SP to Single Prior, KL to 10\% KL$[q||p]$.}
\resizebox{\columnwidth}{!}{
\begin{tabular}{c|cc|cc}
\hline
\multirow{2}{*}{\textbf{Setup}} & \multicolumn{2}{c|}{\textbf{Non-Normative}} & \multicolumn{2}{c}{\textbf{Normative}} \\
\cline{2-3}
\cline{4-5}
 & CER & WER & CER & WER \\
\hline
\multicolumn{1}{l|}{0-shot Inference}  & 40.38 $\pm$ 0.00 & 82.11 $\pm$ 0.00 & \textbf{2.01 $\pm$ 0.00} & 6.18 $\pm$ 0.00 \\
\hline
\multicolumn{1}{l|}{Full Fine-tuning} & 22.60 $\pm$ 1.85 & 46.43 $\pm$ 2.74 & 2.40 $\pm$ 0.34 & 7.83 $\pm$ 0.72 \\
\multicolumn{1}{r|}{+ WD} & 22.53 $\pm$ 1.55 & 46.17 $\pm$ 2.66 & 2.38 $\pm$ 0.27 & 7.66 $\pm$ 0.49 \\
\hline
\multicolumn{1}{l|}{Standard LoRA} & 23.85 $\pm$ 0.51 & 46.64 $\pm$ 1.47 & 2.42 $\pm$ 0.21 & 7.11 $\pm$ 0.40\\
\multicolumn{1}{r|}{+ WD} & 23.11 $\pm$ 0.44 & 46.18 $\pm$ 1.29 & 2.40 $\pm$ 0.19 & 6.98 $\pm$ 0.38\\
\hline
\multicolumn{1}{l|}{MoRA} & 25.87 $\pm$ 0.66 & 49.11 $\pm$ 1.44 & 2.54 $\pm$ 0.15 & 7.80 $\pm$ 0.23 \\
\multicolumn{1}{r|}{+ WD} & 26.43 $\pm$ 0.57 & 48.53 $\pm$ 1.47 & 2.33 $\pm$ 0.14 & 6.97 $\pm$ 0.23 \\
\hline
\multicolumn{1}{l|}{DP VI LoRA + KL} & \textbf{20.09 $\pm$ 0.41} & \textbf{42.86 $\pm$ 1.48} & 2.15 $\pm$ 0.13 & \textbf{6.05 $\pm$ 0.23} \\
\multicolumn{1}{r|}{+ WD} & 31.42 $\pm$ 1.62 & 55.36 $\pm$ 3.51 & 8.21 $\pm$ 0.72 & 16.82 $\pm$ 1.17 \\
\hline
\multicolumn{1}{l|}{SP VI LoRA + KL}& 21.33 $\pm$ 0.51 & 44.85 $\pm$ 1.87 & 2.02 $\pm$ 0.18 & 6.05 $\pm$ 0.27\\
\multicolumn{1}{r|}{+WD} & 26.02 $\pm$ 1.06 & 50.29 $\pm$ 2.09 & 2.33 $\pm$ 0.35 & 7.62 $\pm$ 0.65 \\
\hline
\end{tabular}
}
\label{tab:ErrorBarsModels}
\end{table}

\section{Results and Analysis}
\label{sec:Results}
    
\begin{table}[htb]
\centering
\caption{Results on the UA-Speech (non-normative) and CommonVoice (normative) dataset for different adaptation methods, including a 0-shot baseline, relative to full fine-tuning (= 100\%).}
\label{tab:ua_speech_merged_results}
\footnotesize
\setlength{\tabcolsep}{4pt}
\begin{tabular}{c|c|cc}
\hline
\textbf{Setup} & \textbf{Speech Type} & \textbf{rel. CER} & \textbf{rel. WER} \\
\hline
\multirow{2}{*}{0-shot} & Non-Normative & 271.30\% $\pm$ 0.00\% & 328.80\% $\pm$ 0.00\% \\
 & Normative & \textbf{43.50\% $\pm$ 0.00\%} & \textbf{46.94\% $\pm$ 0.00\%} \\
\hline
\multirow{2}{*}{LoRA} & Non-Normative & 105.32\% $\pm$ 1.71\% & 106.81\% $\pm$ 2.77\% \\
 & Normative & 78.55\% $\pm$ 4.11\% & 81.21\% $\pm$ 4.35\% \\
\hline
\multirow{2}{*}{SP VI LoRA} & Non-Normative & 91.07\% $\pm$ 2.11\% & 91.74\% $\pm$ 2.02\% \\
 & Normative & 44.17\% $\pm$ 5.84\% & 47.29\% $\pm$ 6.11\% \\
\hline
\multirow{2}{*}{DP VI LoRA} & Non-Normative & \textbf{88.94\% $\pm$ 2.36\%} & \textbf{90.24\% $\pm$ 1.78\%} \\
 & Normative & 49.87\% $\pm$ 6.21\% & 55.36\% $\pm$ 5.78\% \\
\hline
\end{tabular}
\end{table}

\begin{table}[ht]
\centering
\footnotesize
\caption{CER and WER on BF-Sprache (100\% $\approx$ 2h) for different sizes of the training set and adaptation methods.}
\resizebox{\linewidth}{!}{
\begin{tabular}{c|cc|cc|cc}
\hline
\multirow{2}{*}{\textbf{Train Data}} & \multicolumn{2}{c|}{\textbf{VI LoRA}} & \multicolumn{2}{c|}{\textbf{Full Fine-tuning}} & \multicolumn{2}{c}{\textbf{LoRA}} \\
\cline{2-7}
 & CER & WER & CER & WER & CER & WER \\
\hline
100\% & \textbf{19.86} & \textbf{42.42} & 22.28 & 48.02 & 23.66 & 47.55  \\
75\%  & \textbf{22.32} & \textbf{44.75} & 24.38 & 49.01 & 25.91 & 51.10 \\
50\%  & \textbf{24.77} & \textbf{50.40} & 28.95 & 66.04 & 28.02 & 58.43 \\
25\%  & \textbf{28.08} & \textbf{56.35} & 33.07 & 70.43 & 31.29 & 66.94  \\
0\%  & 40.38 & 82.11 & 40.38 & 82.11 & 40.38 & 82.11  \\

\hline
\end{tabular}
}
\label{tab:AdaptationVsDataSize}
\end{table}

Table~\ref{tab:ErrorBarsModels} presents a comparative analysis of different adaptation strategies, including standard LoRA, full parameter fine-tuning, and VI LoRA with and without a KL regularization term. All models were fine-tuned on BF-Sprache. The results indicate that VI LoRA, when regularized with a 10\% KL$[q||p]$ term, demonstrates a compelling trade-off. This configuration achieves the lowest CER and WER on the target non-normative speech. Concurrently, it exhibits the least forgetting of normative speech, as evidenced by its leading CER and WER scores, surpassing both standard LoRA and full parameter fine-tuning, which show higher error rates on normative data. These findings generalize across English speakers with varying intelligibility levels in the UA-Speech dataset (Table~\ref{tab:ua_speech_merged_results}). The table reports relative performance differences from full fine-tuning, aggregated over all dysarthric speakers.

We speculate that the performance of VI LoRA with KL divergence regularization stems from its ability to effectively adapt to non-normative speech characteristics while mitigating catastrophic forgetting. The KL divergence term acts as a regularizer, penalizing significant deviations of the adapted LoRA weights ($q$) from the original pre-trained weight distribution ($p$). This constraint likely encourages the model to learn the specific variations of non-normative speech more parsimoniously, preventing overly aggressive updates that could shift the weights too far from their initial state, which is beneficial for normative speech. This allows for controllable adaptation without sacrificing much of the model's generalization ability to the pre-trained normative speech patterns.

Table~\ref{tab:AdaptationVsDataSize} reports CER/WER on non-normative speech in the BF-Sprache dataset across varying training set sizes. The test set is fixed, while training sets are constructed from randomly selected instances of the remaining data. VI LoRA consistently outperforms all baselines, with its advantage being most pronounced when less data is available. While full fine-tuning underperforms even standard LoRA in low-data settings, it surpasses LoRA once the full training set is used. This trend is particularly evident in WER.

As illustrated in Table \ref{tab:combined_examples_metrics_phonetic}, a qualitative analysis of the transcription outputs reveals a critical distinction between the error patterns of full fine-tuning and our proposed VI LoRA model on out-of-distribution (OOD) phrases, specifically targeting semantically rare or hyper-locally relevant terms (e.g., ``Wiedikon'') that lack strong language model priors. The fully fine-tuned model consistently exhibits a form of structured hallucination. For instance, it transcribes the Japanese place name "Higashirinkan" as the grammatically plausible but semantically unrelated German sentence "Ein Gassi rennt da." This suggests the model defaults to familiar linguistic patterns when faced with novel acoustic signals, effectively pattern-matching to the closest structure in its learned distribution.

In stark contrast, VI LoRA produces transcriptions that are phonetically much closer to the ground truth, such as "Higashirenpa." While imperfect, this output demonstrates a failure mode that is grounded in the acoustic evidence rather than learned linguistic priors. This highlights a significant limitation of standard metrics like WER and CER, although both models yield high error rates, the nature of VI LoRA's errors is far more interpretable and useful, as it preserves crucial phonetic information.

We speculate that this robustness stems from the stochastic nature of VI LoRA. The inherent variance across multiple forward passes may disrupt the model's tendency for rigid pattern matching. By marginalizing over these varied predictions, the model is forced to find the smallest common ground, which appears to be the underlying phonetics of the input rather than a pre-learned grammatical structure.

\begin{table}[!htb]
\centering
\caption{Qualitative comparison of transcription examples for out-of-distribution phrases. Phonetic transcriptions (IPA) are provided in \textit{italics} to aid interpretation for non-German speakers.}
\label{tab:combined_examples_metrics_phonetic}
\small 
\resizebox{\columnwidth}{!}{
\begin{tabular}{llc}
\hline
\textbf{System} & \textbf{Transcription Output} & \textbf{PER/CER} \\
\hline
Ground Truth: & ``Wiedikon, Enge, Thalwil, Baar.'' & - \\
& \textit{\textipa{["vi:dIkOn, "EN@, "ta:lvIl, ba:r]}} & - \\
\hline
Full Fine-tuning & ``Wie die kann, eine, teilweise, war.'' & \multirow{2}{*}{56.0 / 45.7} \\
& \textit{\textipa{[vi: di: kan "aIn@ "taIlvaIz@ va:r]}} &  \\
\addlinespace[0.3em] 
\textbf{VI LoRA} (\textbf{ours}) & ``Vidikon, Enne, Talwil, Borg.'' & \multirow{2}{*}{\textbf{20.0} / \textbf{25.0}} \\
& \textit{\textipa{["vi:dIkOn, "En@, "ta:lvIl, bOrk]}} &  \\
\hline

Ground Truth: & ``Higashirinkan.'' & - \\
& \textit{\textipa{[Ci ga Si RiN kaN]}} & - \\
\hline
Full Fine-tuning & ``Ein Gassi rennt da.'' & \multirow{2}{*}{86.7 / 63.2} \\
& \textit{\textipa{[aIn "gasi rEnt da]}} &  \\
\addlinespace[0.3em]
\textbf{VI LoRA} (\textbf{ours}) & ``Higashirenpa.'' & \multirow{2}{*}{\textbf{26.7} / \textbf{25.0}} \\ 
& \textit{\textipa{[Ci ga Si Rempa]}} &  \\
\hline
\end{tabular}
}

\end{table}

\section{Discussion and Conclusions}
\label{sec:Discuss}
In this work, we propose a novel personalization framework using Bayesian LoRA. Our experimental results demonstrate substantial improvements in recognition accuracy for speech-impaired individuals across different intelligibility levels and languages, while maintaining efficient parameter usage. Our experiments revealed several key insights. Our dual prior approach to VI LoRA (Table \ref{tab:ErrorBarsModels} for BF-Sprache and Table \ref{tab:ua_speech_merged_results} for UA-Speech) showed particularly strong performance, reducing non-normative speech CER to 20.09\% (compared to 21.33\% for single prior) while maintaining reasonable normative speech performance. This suggests that modeling the bimodal distribution of pre-trained weights significantly improves adaptation capability. The cross-lingual results are especially encouraging, with our framework generalizing effectively from English to German despite the latter's phonological complexity. Furthermore, the observed trade-off between performance on non-normative and normative speech highlights a key design consideration: systems optimized solely for impaired speech may sacrifice generalization, suggesting that multi-objective training strategies could further improve generalizability. While promising, our work has some limitations. Our pipeline assumes independent factorization of $q_\phi(A,B)$ by disentangling it as $q_{\phi_A}(A)q_{\phi_B}(B)$. While computationally efficient, it may not best capture the interactions between LoRA adapter matrices \citep{zhu2024asymmetry, liu2024dora}. The main limitation remains the small speaker pool in the BF-Sprache dataset, due to previous resource and ethical approval constraints. With the ethical approval now secured, our immediate future work will focus on expanding this dataset by recruiting a larger, more diverse group of speakers across various conditions and intelligibility levels. Our further work will focus on expanding the speaker base in BF-Sprache over different conditions and intelligibility levels as well as incorporating VI LoRA in an active learning setting for continuous speaker-specific adaptation. 

\newpage


\clearpage
\footnotesize
\bibliographystyle{IEEEbib}
\bibliography{mybib.bib}





\end{document}